\documentclass[conference]{IEEEtran}
\IEEEoverridecommandlockouts
\usepackage{cite}
\usepackage{amsmath,amssymb,amsfonts}
\usepackage{algorithmic}
\usepackage{graphicx}
\usepackage{textcomp}
\usepackage{xcolor}
\usepackage[caption=false,font=footnotesize]{subfig}
\def\BibTeX{{\rm B\kern-.05em{\sc i\kern-.025em b}\kern-.08em
    T\kern-.1667em\lower.7ex\hbox{E}\kern-.125emX}}
\DeclareUnicodeCharacter{FB01}{fi}
\begin{document}

\title{Near-Field Localization with an Exact Propagation Model in Presence of Mutual Coupling
\thanks{This work was supported by Leverhulme Trust under Research Leadership Award under Grant RL-2019-019 and British Telecom under UKRI Future Leaders Fellowship grant MR/T019980/1.}
}

\author{
	\IEEEauthorblockN{
		Zohreh Ebadi\IEEEauthorrefmark{1},
		Amir Masoud Molaei\IEEEauthorrefmark{1},
		Muhammad Ali Babar Abbasi\IEEEauthorrefmark{1},
		Simon Cotton\IEEEauthorrefmark{1},
        Anvar Tukmanov\IEEEauthorrefmark{2}, \\
        and Okan Yurduseven\IEEEauthorrefmark{1}
	}
	\IEEEauthorblockA{
				\IEEEauthorrefmark{1} CWI, EEECS, Queen's University Belfast, BT3~9DT Belfast, U.K.
			} 
	\IEEEauthorblockA{
				\IEEEauthorrefmark{2} BT~Labs,~Adastral~Park,~BT~Group,~Ipswich,~U.K.
			}
	Email: zebadi01@qub.ac.uk
}

\maketitle

\begin{abstract}
Localizing near-field sources considering practical arrays is important in wireless communications. Array-based apertures exhibit mutual coupling between the array elements, which can significantly degrade the performance of the localization method. In this paper, we propose two methods to localize near-field sources by direction of arrival (DOA) and range estimations in the presence of mutual coupling. The first method utilizes a two-dimensional search to estimate DOA and the range of the source. Therefore, it suffers from a high computational load. The second method reduces the two-dimensional search to one-dimensional, thus decreasing the computational complexity while offering similar DOA and range estimation performance. Besides, our second method reduces computational time by over 50\% compared to the multiple
signal classification (MUSIC) algorithm.
 
\end{abstract}

\begin{IEEEkeywords}
DOA and range estimations, exact model, mutual coupling, near-field localization
\end{IEEEkeywords}

\section{introduction}
Source localization is crucial in sonar, radar, and wireless communication applications~\cite {Krim1996,Molaei2022}. Conventionally, localization algorithms can be classified based on source location; far-field and near-field localization algorithms~\cite{Zhang2023,Zhang2020}. If the source is in the near-field region, known as the Fresnel region, then the far-field localization algorithms will not be able to estimate the location of the source. This is due to the fact that the received wavefront from the far-field source becomes planar. Therefore, the far-field wavefront will be characterized by only the direction of arrival (DOA), while both DOA and range characterize the near-field wavefront because the received signal from the near-field source has a spherical shape\cite{Zuo2019,Huang1991}. 

Mutual coupling refers to the electromagnetic interaction between the antenna elements of an array. This interaction can reduce the accuracy of the localization algorithm \cite{Chen2019,Lan2023,Liu2023}. Based on the type of elements in the antenna array, mutual coupling can be categorized as either direction-independent or direction-dependent \cite{Ge2020,Qi2019,Elbir2017}. For arrays consisting of omnidirectional antennas, the mutual coupling can be considered direction-independent. However, for practical arrays with non-omnidirectional antennas, the mutual coupling is direction-dependent \cite{Elbir2017,Qi2019}. 

Numerous algorithms have been developed over the decades to estimate the location of a source in the presence of mutual coupling~\cite{Qi2019,Zhen2018,Friedlander1991,Famoriji2021,Abedin2012,Xie2016,Wen2023,Chen2019}. These algorithms can be categorized into active calibration~\cite{Qi2019} and self-calibration\cite{Friedlander1991}. The active calibration algorithms rely on prior knowledge of the locations of the sources. However, the accuracy of these methods significantly reduces in the presence of localization errors~\cite{Qi2019}. On the other hand, the self-calibration methods estimate the mutual coupling and location of the source jointly. For example, the method in \cite{Friedlander1991} is an iterative algorithm that estimates DOA and mutual coupling. However, these methods only consider the far-field sources.

For near-field DOA and range estimations in the presence of mutual coupling, methods such as those presented in \cite{Famoriji2021,Abedin2012,Xie2016,Wen2023,Chen2019} were proposed. Methods presented in \cite{Famoriji2021,Abedin2012} are developed for direction-independent mutual coupling. On the other hand, the methods in~\cite{Xie2016,Wen2023,Chen2019} consider mixed far-field and near-field sources. In the mixed sources scenarios, the far-field sources' DOA and mutual coupling are first estimated. Then, the estimated mutual coupling is used to compensate for the received signal. This compensated received signal is then used to predict the near-field source location by estimating the DOA and the range of the near-field sources. However, the mutual coupling, DOA, and range must all be estimated simultaneously in scenarios where only the near-field sources are considered. It will cause a three-dimensional search algorithm to estimate the three mentioned unknowns, which can drastically increase the computational complexity.

In this work, we consider near-field source localization in the presence of the direction-dependent mutual coupling between the receiver array elements. We use the exact system model~\cite{Chen2023,Friedlander2019} for the near-field sources because using the approximated model~\cite{Li2021,Cheng2022} causes an error in DOA and range estimations. This is due to the fact that the exact model, unlike the approximated model, relies on the behavior of electromagnetic waves in the near-field~\cite{Friedlander2019}. We propose two methods to estimate the location of the source in the near-field. The first method is inspired by the one described in \cite{Bazzi2016}. The main difference between our first method and method in \cite{Bazzi2016} is that the method in \cite{Bazzi2016} is designed for the far-field source location estimation, while our method is for estimating source location in the near-field. Since our first method relies on two-dimensional searches, we propose a second method, which employs one-dimensional searches. Therefore, the main contribution of this work is presenting two algorithms for near-field source localization in the presence of direction-dependent mutual coupling. In addition, we reduce the search to two-dimensional and one-dimensional problems, minimizing the computational complexity.

The rest of the paper is organized as follows. In Section \ref{System_model_Section}, we present the system model. Section \ref{DOAandRangeEstimation} explains the multiple signal classification (MUSIC) algorithm for the localization of the source in the near-field. Section \ref{proposed_algorithms} presents our methods for near-field source localization. Section \ref{Simulation} provides simulation results. Finally, conclusions are drawn in Section \ref{Conclusion}.

\section{system model}\label{System_model_Section}
\begin{figure}[!t]
\centerline{\includegraphics[width=3.2in]{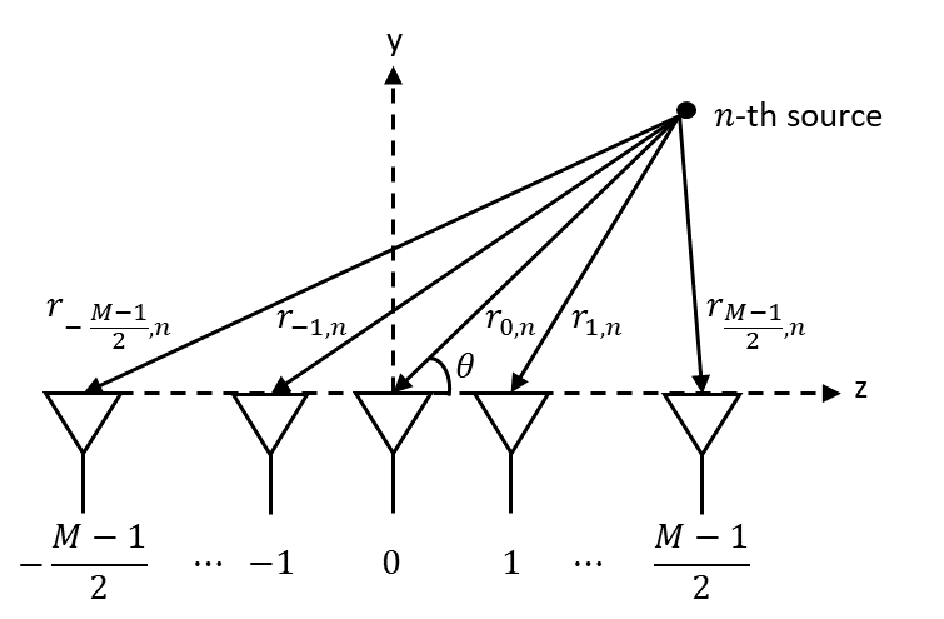}}
\caption{The considered system model with the sources in the near-field.\hspace{-10pt}}
\label{sourcelocation}
\end{figure}

Suppose that $N$ narrowband signals impinging on a uniform linear array (ULA), as shown in Fig.~\ref{sourcelocation}. 

The ULA consists of $M$ elements, where $M$ is odd. The inter-element spacing of elements is $d={\lambda}/{2}$, where $\lambda$ is the signal wavelength. The output signal of the ULA can be expressed as~\cite{Ge2020}
\begin{align}\label{system_model}
   \mathbf{Y} = \mathbf{\Tilde{A}}(\boldsymbol{\theta},\mathbf{r})\mathbf{S}+\mathbf{W},
\end{align}
where $\mathbf{S}=[s_1(t)\ s_2(t)\ \cdots \ s_N(t)]^T$ represents the vector of the transmitted non-coherent signals, $\mathbf{W}=[w_1(t)\ w_2(t)\ \cdots \ w_M(t)]^T$ is white Gaussian noise vector with a zero mean and variance $\sigma^2$, and $t=1,2,\cdots,L$. $L$ is the number of the snapshots. $\boldsymbol{\theta} =[\theta_{1}\ \theta_{2}\ \cdots\ \theta_{N}]$ is the DOA vector of the received signals at the reference element, i.e., zeroth element in the ULA, and $\mathbf{r}=[r_{0,1}\ r_{0,2}\ \cdots\ r_{0,N}]$ is the vector of range of sources where the range is defined as the distance between the $n$-th source and the reference element, where $n=1, 2,\cdots, N$. We assume that all sources are in the near-field (Fresnel) region of the antenna array, $r_{0,n}\in(0.62(D^3/\lambda)^\frac{1}{2},2D^2/\lambda)$, where $D$ is array aperture~\cite{Zuo2019,Molaei2022}. Besides, $\mathbf{\Tilde{A}}(\boldsymbol{\theta},\mathbf{r})=[\mathbf{\Tilde{a}}(\theta_1,r_{0,1})\  \mathbf{\Tilde{a}}(\theta_2,r_{0,2})\  \cdots \ \mathbf{\Tilde{a}}(\theta_N,r_{0,N})]$ is $M\times N$ matrix. Each vector in $\mathbf{\Tilde{A}}(\boldsymbol{\theta},\mathbf{r})$ can be written as~\cite{Liao2012}
\begin{align}\label{a_Tilda}
    \mathbf{\Tilde{a}}(\theta_n,r_{0,n})=\mathbf{C}(\theta_n)\mathbf{a}(\theta_n,r_{0,n}),
\end{align}
where $\mathbf{C}(\theta_n)\in \mathbb{C}^{M\times M}$ is a direction-dependent mutual coupling matrix of ULA at the $n$-th angle as follows~\cite{Ge2020,Elbir2017}:
\begin{align}\label{MutualCouplingMatrix}
    \mathbf{C}(\theta_n)=\text{Toeplitz}\{\mathbf{z}(\theta_n)\},
\end{align}
where $\mathbf{z}=[\mathbf{c}^T(\theta_n),\mathbf{0}]^T$. $\mathbf{c}(\theta_n)=[c_1(\theta_n),c_2(\theta_n),\cdots,c_P(\theta_n)]^T$ is mutual coupling coefﬁcients vector with dimension of $P\times1$, and $P<M$. This is because the mutual coupling coefficients decrease as the distance between elements increases. Beyond $P$ inter-element spacings, coupling coefficients become negligible. Furthermore, $\mathbf{0}$ is vector with dimension of $1\times(M-P)$. Besides,
$\mathbf{a}(\theta_n,r_{0,n})$ is $M\times 1$ steering vector, where each element in the steering vector based on the exact model is given by~\cite{Friedlander2019}
\begin{align}\label{steering_vector1}
    \mathbf{a}(\theta_{n},r_{0,n}) = \frac{r_{0,n}}{r_{m,n}}e^{-j\tau_{m,n}},
\end{align}
where $m=-\frac{M-1}{2},\ \cdots,\ -1,\ 0, \ 1,\ \cdots,\ \frac{M-1}{2}$. Furthermore, $\tau_{m,n}$ is the phase difference between the $n$-th source and $m$-th element in the array as follows:
\begin{align}\label{Tau}
    \tau_{m,n}= \frac{2\pi}{\lambda}(r_{m,n}-r_{0,n}),
\end{align}
where $r_{m,n}$ is the distance between the $n$-th source and the array's $m$-th element and is given by~\cite{Friedlander2019}
\begin{align}\label{distance}
    r_{m,n} = \sqrt{r^2_{0,n}+m^2d^2-2mdr_{0,n}\cos{\theta_n}}~.
\end{align}

In most literature~\cite{Zhang2018, Molaei2022,Guanghui2020}, \eqref{steering_vector1} is simplified using the Fresnel approximation and neglecting the magnitude as follows:
\begin{align}\label{steering_vector_nearfield}
    \mathbf{a}^{'}(\theta_{n},r_{0,n}) = e^{j(\gamma_nm+\eta_n m^2)},
\end{align}
where
\begin{align}\begin{split}
    &\gamma_n=-\frac{2\pi d}{\lambda}\cos{\theta_n}, \\
    &\eta_n=\frac{\pi d^2}{\lambda r_{0,n}}\sin^2{\theta_n}.
    \end{split}
\end{align}
The model in \eqref{steering_vector_nearfield} is the approximated model.

With the array output signal \eqref{system_model}, this paper aims to estimate $\boldsymbol{\theta}$ and $\mathbf{r}$ of non-coherent sources in the presence of unknown
mutual coupling.

\section{doa and range estimation}\label{DOAandRangeEstimation}
One standard way to estimate the DOA and range is using the MUSIC algorithm~\cite{Schmidt1986}. To do this, the covariance matrix of the received signal can be written as
\begin{align}\label{covariance}
    \mathbf{R}_y=\frac{1}{L}\sum_{t=1}^{L}\mathbf{y}(t)\mathbf{y}^H(t).
\end{align}
where $\mathbf{y}(t)$ is $M\times1$ vector and denotes the received signal at $t$-th snapshot. The eigendecomposition of \eqref{covariance} is given by
\begin{align}\label{eigendecomposition}
   \mathbf{R}_y=\mathbf{U}\mathbf{\Sigma}\mathbf{U}^H. 
\end{align}
where $\boldsymbol{\Sigma}$ is eigenvalues of $R_y$ and $\mathbf{U}=[\mathbf{U}_s\ \mathbf{U}_w]$ is eigenvector matrix. $\mathbf{U}_s\in \mathbb{C}^{M\times N}$ and $\mathbf{U}_w\in \mathbb{C}^{M\times (M-N)}$ denote signal and noise subspaces, respectively. Then, the unknowns can be jointly obtained by searching $N$ peaks of the MUSIC spatial spectrum as follows:
\begin{align}\label{MUSIC_Spectrum}
    P(\theta_n,r_{0,n})=\frac{1}{\textbf{a}^H(\theta_n,r_{0,n})\mathbf{C}^H(\theta_n)\textbf{U}_w\textbf{U}_w^H\mathbf{C}(\theta_n)\textbf{a}(\theta_n,r_{0,n})}.
\end{align}
However, \eqref{MUSIC_Spectrum} cannot be used to estimate the unknowns including $\mathbf{C}(\theta_n)$, DOA, and range. Due to the complex values in $\mathbf{C}(\theta_n)$, defining the search area for these values may be infeasible. On the other hand, $\theta_n\in [-\frac{\pi}{2},\frac{\pi}{2}]$ and $r_{0,n}\in(0.62(D^3/\lambda)^\frac{1}{2},2D^2/\lambda)$. Thus, defining the search area for $\theta_n$ and $r_{0,n}$ in \eqref{MUSIC_Spectrum} is feasible.

\section{proposed algorithms}\label{proposed_algorithms}
In this section, we propose two methods to estimate the DOA and the range in the presence of mutual coupling.
\subsection{First Algorithm}
In this subsection, we first examine whether the transformation matrix, commonly used in the literature, can be applied in near-field scenarios. To do this, according to \eqref{a_Tilda}, $\mathbf{\Tilde{a}}(\theta_n,r_{0,n})$ can be written as follows:
\begin{align} 
    \label{Transform_defenition}
&\mathbf{\Tilde{a}}(\theta_n,r_{0,n})= \nonumber \\
    &\sum_{m=1}^{P}\mathbf{E}_m\mathbf{a}(\theta_n,r_{0,n})c_m(\theta_n)= \nonumber \\
    &[\mathbf{E}_1\mathbf{a}(\theta_n,r_{0,n})\ \mathbf{E}_2\mathbf{a}(\theta_n,r_{0,n})\ \cdots \ \mathbf{E}_{P}\mathbf{a}(\theta_n,r_{0,n})] \nonumber \\
    &[c_1(\theta_n)\ c_2(\theta_n)\ \cdots \ c_P(\theta_n)]^T= \nonumber \\
    &\mathbf{X}(\theta_n,r_{0,n})\mathbf{c}(\theta_n),
\end{align}
where $\mathbf{X}(\theta_n,r_{0,n})$ is a ${M\times P}$ matrix. Besides, $\mathbf{E}_m$ is defined as
\begin{align}
    [\mathbf{E}_m]_{ij}=
    \begin{cases}
      1, & [\mathbf{C}]_{ij}(\theta_n)=z_m(\theta_n)\\
      0, & \text{otherwise}.
    \end{cases} 
\end{align}
where $[\cdot]_{ij}$ indicates ${ij}$-th element of the corresponding matrix. Note that the transformation matrix in \eqref{Transform_defenition} differs from that in \cite{Bazzi2016,Ge2020,Elbir2017a,Elbir2017} because $\mathbf{X}$ in \eqref{Transform_defenition} is a function of both $\theta$ and $r$, while in \cite{Bazzi2016,Ge2020,Elbir2017a,Elbir2017}, $\mathbf{X}$ is only a function of $\theta$.

By inserting \eqref{Transform_defenition} into \eqref{MUSIC_Spectrum}, \eqref{MUSIC_Spectrum} can be written as
\begin{align}\label{MUSIC_Spectrum3}
    \{\hat{\theta}_n,\hat{r}_{0,n}\}_{n=1}^N=\arg\max_{\theta_n,r_{0,n},\mathbf{c}}\frac{1}{\textbf{c}^H(\theta_n)\boldsymbol{\Omega}(\theta_n,r_{0,n})\textbf{c}(\theta_n)}.    
\end{align}
where $\boldsymbol{\Omega}(\theta_n,r_{0,n})=\mathbf{X}^H(\theta_n,r_{0,n})\textbf{U}_w\textbf{U}_w^H\mathbf{X}(\theta_n,r_{0,n})$. To find the unknowns in \eqref{MUSIC_Spectrum3}, we solve the following optimization problem for given $\theta$ and $r$:
\begin{align}\label{Optimization}
    &\min_{\mathbf{c}}\textbf{c}^H(\theta_n)\boldsymbol{\Omega}(\theta_n,r_{0,n})\textbf{c}(\theta_n) \nonumber\\ 
    &\text{subject to}\  \mathbf{e}_1^H\mathbf{c}(\theta_n)=1.    
\end{align}
where $\mathbf{e}_1$ is the first column of the unitary matrix. The detailed solution of \eqref{Optimization} is provided in \cite{Bazzi2016}. To solve this optimization problem, we form the Lagrangian function as follows
\begin{align}\label{lagrangian}
    \mathcal{L}(\mathbf{c},\beta)=\mathbf{c}^H(\theta_n)\boldsymbol{\Omega}(\theta_n,r_{0,n})\mathbf{c}(\theta_n)-\beta(\mathbf{e}_1^H\mathbf{c}(\theta_n)-1)
\end{align}
where $\beta$ is Lagrange multiplier. By setting $\frac{\partial \mathcal{L}(\mathbf{c},\beta)}{\partial \mathbf{c}(\theta_n)}$ (first derivative of $\mathcal{L}(\mathbf{c},\beta)$ with respect to $\mathbf{c}(\theta_n)$) to zero and using the constraint $\mathbf{e}_1^H\mathbf{c}(\theta_n)=1$, the estimated DOA and range are given by
\begin{align}\label{Algorithm1_Solution}
    \{\hat{\theta}_n,\hat{r}_{0,n}\}_{n=1}^N=\arg\max_{\theta_n,r_{0,n}}\textbf{e}^H_1\boldsymbol{\Omega}^{-1}(\theta_n,r_{0,n})\textbf{e}_1.    
\end{align}
However, this method requires a two-dimensional search. Therefore, the computational complexity of this method is high. In the next method, we propose a second algorithm with less computational complexity than the first.

\subsection{Second Algorithm}
This method also utilizes the exact model of the steering vector. The algorithm steps are as follows:

1) Initialize: Since this method needs an initial value for one of the unknowns, we use the approximated model to find the initial value for DOA. $\mathbf{C}(\theta_n)\mathbf{a}^{'}(\theta_n,r_{0,n})$ can be written as follows
\begin{align} \label{Transformation_Approx}
    &\mathbf{C}(\theta_n)\mathbf{a}^{'}(\theta_n,r_{0,n})= \nonumber \\
        &\begin{bmatrix}
            g_1 & c_1g_2 & c_2g_3 & \cdots & c_{M-1}g_M\\
            c_1g_1 & g_2 & c_1g_3 & \cdots & c_{M-2}g_M\\
            \vdots & \vdots & \vdots & \vdots & \vdots\\
            c_{M-1}g_1 & c_{M-2}g_2 & c_{M-3}g_3 & \cdots & g_M
        \end{bmatrix}\begin{bmatrix}
            v_1\\
            v_2\\
            \vdots\\
            v_M
        \end{bmatrix}= \nonumber \\
        &\mathbf{B}(\theta_n)\mathbf{v}(\theta_n,r_{0,n}),
\end{align}
where
\begin{align} \begin{split}
    \label{approximateModel2_elements}
    {v}_m(\theta_n,r_{0,n})=e^{j\eta_n m^2} \\
    {g}_m(\theta_n)=e^{j\gamma_nm}.     
    \end{split}
\end{align}

Therefore, 
by inserting \eqref{Transformation_Approx} into \eqref{MUSIC_Spectrum}, \eqref{MUSIC_Spectrum} can be written as
\begin{align}\label{Algorithm1_Solution2}
&\{\hat{\theta}_n,\hat{r}_{0,n}\}_{n=1}^N= \nonumber \\
&\arg\min_{\theta_n,r_{0,n}}\textbf{v}^H(\theta_n,r_{0,n})\mathbf{B}^H(\theta_n)\textbf{U}_w\textbf{U}^H_w\mathbf{B}(\theta_n)\textbf{v}^H(\theta_n,r_{0,n}).    
\end{align}
To estimate the initial value for $\theta_n$, \eqref{Algorithm1_Solution2} can be approximated by
\begin{align}\label{Algorithm1_Solution3}
\{\hat{\theta}_n\}_{n=1}^N=\arg\min_{\theta_n}\mathbf{B}^H(\theta_n)\textbf{U}_w\textbf{U}^H_w\mathbf{B}(\theta_n).    
\end{align}
Using the same procedure as \eqref{Transform_defenition}, $\mathbf{B}(\theta_n)=\mathbf{X}(\theta_n)\mathbf{c}(\theta_n)$. Therefore, \eqref{Algorithm1_Solution3} can be written as follows
\begin{align}\label{Algorithm1_Solution4}
\{\hat{\theta}_n\}_{n=1}^N=
\arg\min_{\theta_n}\textbf{c}^H(\theta_n)\mathbf{X}^{H}(\theta_n)\textbf{U}_w\textbf{U}^H_w\mathbf{X}(\theta_n)\textbf{c}(\theta_n).    
\end{align}
Therefore, to estimate DOA, the following optimization problem needs to be solved for given $\theta_n$
\begin{align}\label{optimization4}
    &\min \mathbf{c}^H(\theta_n)\boldsymbol{\Omega}^{'}(\theta_n)\mathbf{c}(\theta_n)\nonumber\\
    &\text{subject to}\ \mathbf{e}^H_1\mathbf{c}(\theta_n)=1.
\end{align}
where $\boldsymbol{\Omega}^{'}(\theta_n)=\mathbf{X}^{H}(\theta_n)\textbf{U}_w\textbf{U}^H_w\mathbf{X}(\theta_n)$. Therefore, the initial value of DOA is as follows
\begin{align}\label{Initial_DOA}
    \hat{\theta}^{(i)}_n=\arg\max_{\theta_n}\mathbf{e}^H_1{\boldsymbol{\Omega}^{'}}^{-1}(\theta_n)\mathbf{e}_1.
\end{align}
where $i$ denotes the iteration index.

2) Estimate the range of the source: $\hat{r}_{0,n}$ can be found using 
\begin{align}\label{Initial_r}
    \hat{r}_{0,n}=\arg\max_{r_{0,n}}\mathbf{e}^H_1\boldsymbol{\Omega}^{-1}(\hat{\theta}_n,r_{0,n})\mathbf{e}_1.
\end{align}

3) Estimate the DOA of the source: since the initial estimate $\hat{\theta}_n$ in \eqref{Initial_DOA} relies on an approximated model and certain assumptions, the resulting $\hat{r}_{0,n}$ can be utilized to modify the DOA.
\begin{align}\label{Estimated_DOA}
    \hat{\theta}^{(i+1)}_n=\arg\max_{\theta_n}\mathbf{e}^H_1\boldsymbol{\Omega}^{-1}(\theta_n,\hat{r}_{0,n})\mathbf{e}_1.
\end{align}

4) Check the constraint: If $|\hat{\theta}^{i+1}_n-\hat{\theta}^{i}_n|<q$, then terminate; otherwise, return step 2. In the simulations, $q$ is a small positive number.

\begin{figure}[!t]
    \centering
    \includegraphics[width=2.2in]{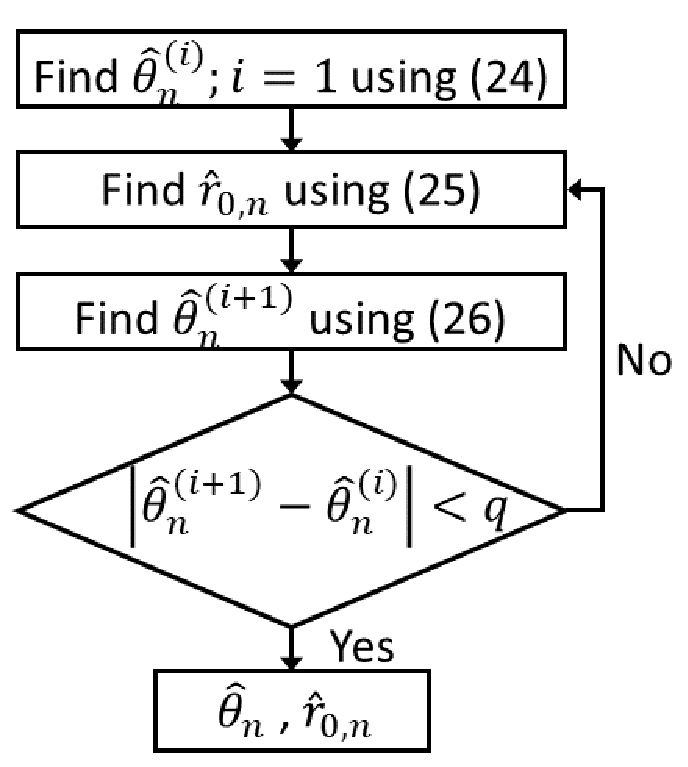}
    \caption{Second algorithm for DOA and range estimation of sources in the near-field.\hspace{-10pt}}
    \label{SecondAlgorithm}
\end{figure}
Based on the above steps, the second algorithm is
summarized in Fig.~\ref{SecondAlgorithm}. This method requires one-dimensional searches to estimate the DOA and the range of the near-field source.

\section{simulation results}\label{Simulation}
In this section, simulations are conducted to validate the performance of the proposed methods. In these simulations, we consider a ULA consisting of five elements operating at a frequency of $5$~GHz. The number of snapshots equals $200$. We consider one source in different locations within the Fresnel region, $[1.75\lambda,8\lambda]$, with DOA of $[30^\circ,40^\circ,50^\circ,60^\circ]$ and range of $3.3\lambda$. We assume that the mutual coupling is direction-dependent. So, for each DOA, the mutual coupling coefficients are generated randomly \cite{Elbir2017}. The number of nonzero mutual coupling coefficients, $P$, is considered to be equal to $3$. All simulation results are obtained based on the average of $K=1000$ independent trials. As a performance metric for the accuracy of the estimations, the root mean square error (RMSE) is used. RMSE is defined as~\cite{Molaei2022}
\begin{align}
    \text{RMSE}_\alpha=\sqrt{\frac{1}{KN}\sum_{k=1}^{K}\sum_{n=1}^{N}(\hat{\alpha}_{n,k}-\alpha_{n})^2}.
\end{align}
where $\alpha$ and $\hat{\alpha}$ denote the true and estimated values of DOA or range, respectively. In all simulations, $q$ for the second method is $0.1^\circ$.

\begin{figure}[!t]
\subfloat[]{\includegraphics[width=\columnwidth]{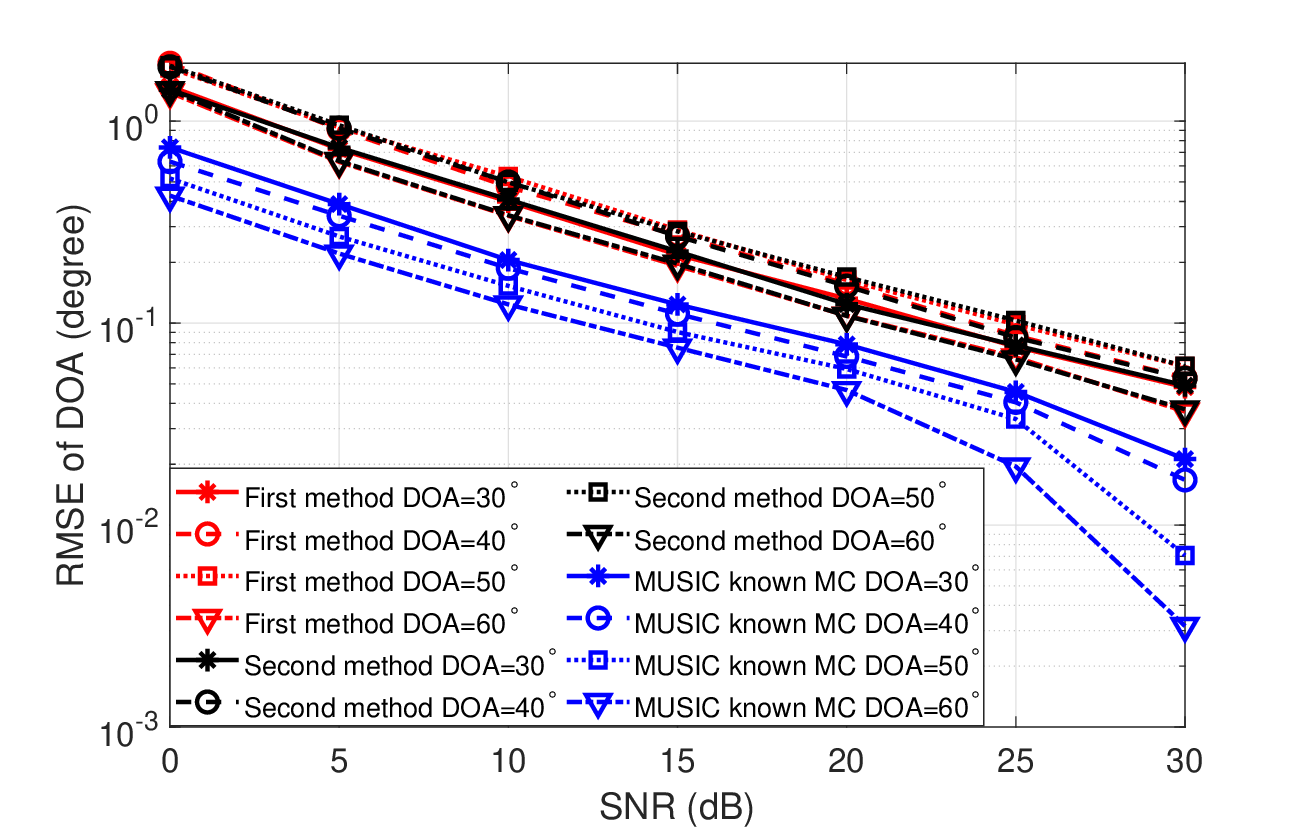}%
\label{RMSEDOAVsSNR}}
\vfil
\subfloat[]{\includegraphics[width=\columnwidth]{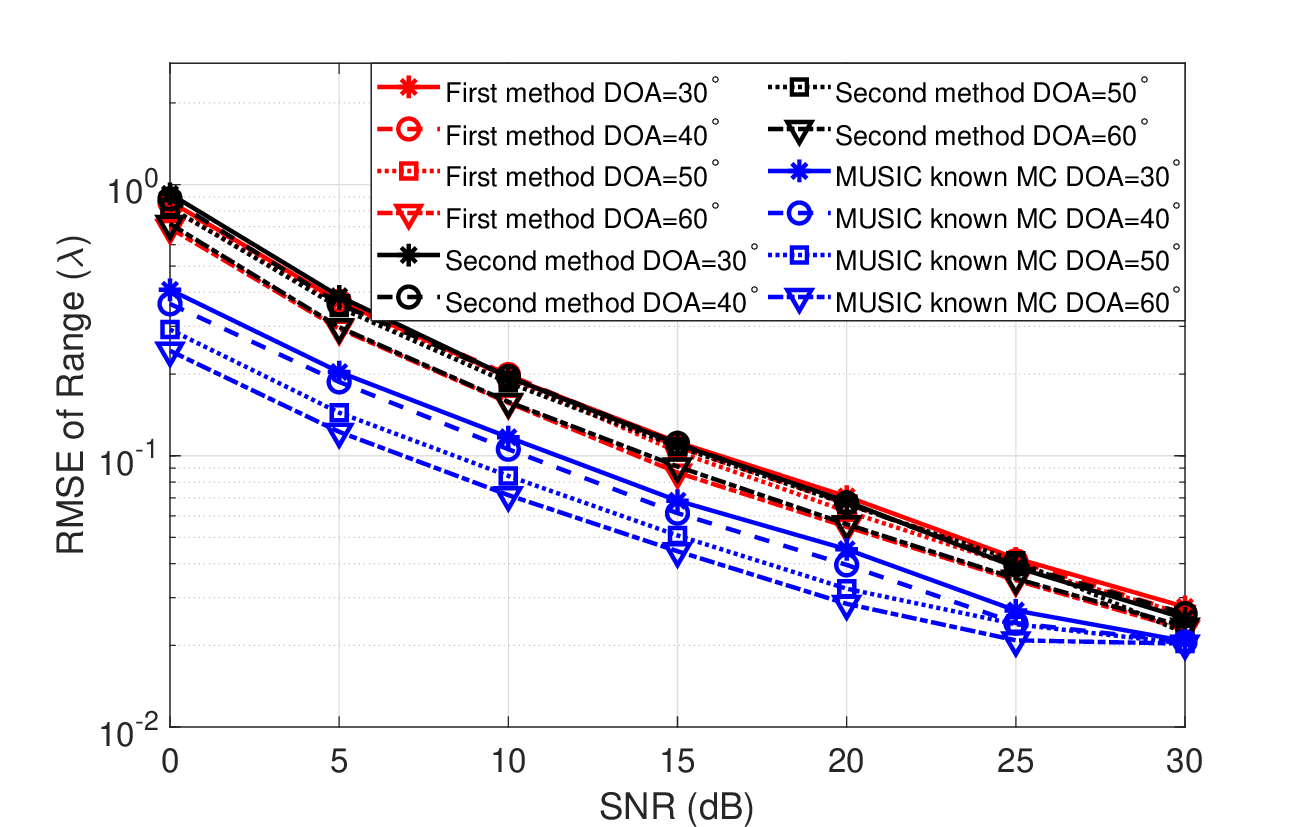}%
\label{RMSERangeVsSNR}}
\caption{RMSE of estimated (a) DOA, (b) range versus SNR.\hspace{-10pt}}
\label{RMSEUnknownsVSSNR}
\end{figure}

\textbf{Example 1:} This simulation investigates the RMSE of DOA and range versus signal-to-noise ratio (SNR), as shown in Fig.~\ref{RMSEUnknownsVSSNR}. The SNR varies from $0$~dB to $30$~dB in steps of $5$~dB. The MUSIC algorithm is used as a baseline to compare the accuracy of the proposed methods with it. The mutual coupling is assumed to be known for the MUSIC algorithm. 

As seen in Fig.~\ref{RMSEUnknownsVSSNR}(a), the first and second algorithms' accuracy is almost the same for all locations. Furthermore, the RMSE of the estimated DOA using MUSIC is less than the RMSE of the estimated DOA using the first and second methods. For example, for a DOA of $30^\circ$, the RMSE of the estimated DOA using MUSIC is $0.69^\circ$ less than the RMSE of the estimated DOA using the second method in SNR of $0$~dB. This difference decreases as SNR increases, such that at an SNR of $30$~dB, the difference reaches $0.03^\circ$.

Fig.~\ref{RMSEUnknownsVSSNR}(b) shows that the RMSE of the estimated range using the MUSIC algorithm is lower than the RMSE of the estimated range using the first and second methods for all locations. For instance, for DOA of $30^\circ$, the RMSE of the estimated range using MUSIC is $0.46\lambda$ and $0.01\lambda$ less than the RMSE of the estimated range using the second method in SNR of $0$~dB and $30$~dB, respectively. 

It should be noted that the MUSIC algorithm outperforms the proposed methods since it assumes the knowledge of mutual coupling.

\textbf{Example 2:} In this simulation, the computational load of the methods is compared using the computational time. In this simulation, SNR and $K$ are $10$~dB and $1$, respectively. The simulations are conducted on the same machine with Intel(R) Core(TM) i7-1265U processor and $16$~GB of RAM. MATLAB version is R2022b.

The computational time for the first method, second method, and MUSIC algorithm are $0.90$~sec, $0.12$~sec, and $0.30$~sec, respectively. The second method's computational time is less than the other two methods because it estimates DOA and the range of sources using one-dimensional searches.

\section{conclusion}\label{Conclusion}
In this paper, we proposed two methods to estimate the DOA and range of a source in the near-field, in the presence of mutual coupling.  The first method is based on a two-dimensional search that simultaneously estimates the DOA and range of the source. On the other hand, the second algorithm is an iterative method based on a one-dimensional search. In the latter method, each unknown is estimated separately using the other estimated unknown from the previous step. 

The significance of this work is twofold. Firstly, our proposed methods estimate DOA and the range of the source in near-field without estimating mutual coupling coefficients. As a result, our methods reduce the computational load by avoiding the need to search through a large search area to estimate mutual coupling. Secondly, our second method yields performance similar to our first method while having a lower computational load compared to the first method. In addition, our second method reduces the computational time by more than $50\%$ in comparison to the MUSIC algorithm.

\bibliographystyle{IEEEtran}
\bibliography{IEEEabrv,BibliographyForMutualCoupling}
\vspace{12pt}

\end{document}